\newif\ifarxiv
\arxivtrue  

\ifarxiv
  \documentclass[sigconf,nonacm]{acmart}
\else
  \documentclass[sigconf,anonymous,review]{acmart}
\fi
\newif\ifanon
\ifarxiv\anonfalse\else\anontrue\fi

\usepackage{array}
\usepackage{booktabs}
\usepackage{graphicx}
\usepackage{microtype}
\newcommand{\TopoContainedEW}{37.8\% [36.9, 38.7]}
\newcommand{\TopoContainedUW}{34.1\% [33.5, 34.7]}
\newcommand{\TopoOpenerEW}{9.6\% [9.4, 9.8]}
\newcommand{\TopoOpenerUW}{10.5\% [10.2, 10.9]}
\newcommand{\TopoCloserEW}{15.3\% [15.0, 15.7]}
\newcommand{\TopoCloserUW}{18.3\% [17.8, 18.7]}
\newcommand{\TopoBridgeEW}{37.3\% [36.5, 38.2]}
\newcommand{\TopoBridgeUW}{37.1\% [36.5, 37.7]}
\newcommand{\TopoContainedSingleExchangeEW}{17.1\% [16.6, 17.5]}
\newcommand{\TopoContainedSingleExchangeUW}{19.0\% [18.5, 19.4]}
\newcommand{\TopoContainedMultiExchangeEW}{8.7\% [8.4, 9.1]}
\newcommand{\TopoContainedMultiExchangeUW}{7.4\% [7.1, 7.7]}
\newcommand{\TopoContainedMultipleSessionsEW}{12.0\% [11.4, 12.6]}
\newcommand{\TopoContainedMultipleSessionsUW}{7.7\% [7.4, 8.0]}
\newcommand{\WebBeforeEW}{47.3\% [46.5, 48.2]}
\newcommand{\WebBeforeUW}{51.9\% [51.3, 52.5]}
\newcommand{\WebAfterEW}{39.8\% [39.1, 40.4]}
\newcommand{\WebAfterUW}{42.7\% [42.2, 43.3]}
\newcommand{\ContentBeforeEW}{42.6\% [41.8, 43.3]}
\newcommand{\ContentBeforeUW}{45.5\% [44.9, 46.1]}
\newcommand{\ContentAfterEW}{36.5\% [35.8, 37.3]}
\newcommand{\ContentAfterUW}{38.9\% [38.4, 39.5]}
\newcommand{\SearchAnywhereEW}{42.1\% [41.3, 43.0]}
\newcommand{\SearchAnywhereUW}{46.0\% [45.4, 46.6]}
\newcommand{\SearchBeforeEW}{28.1\% [27.4, 28.7]}
\newcommand{\SearchBeforeUW}{33.8\% [33.2, 34.4]}
\newcommand{\SearchAfterEW}{21.7\% [21.2, 22.2]}
\newcommand{\SearchAfterUW}{24.4\% [23.9, 24.8]}
\newcommand{\SearchBetweenEW}{12.1\% [11.7, 12.6]}
\newcommand{\SearchBetweenUW}{8.7\% [8.4, 9.0]}
\newcommand{\ContainedMultipleResponse}{54.5\% [53.7, 55.3]}
\newcommand{\ContainedSingleExchangeWithinContained}{45.2\%}
\newcommand{\ContainedMultipleAssistantSession}{31.7\% [30.7, 32.7]}
\newcommand{\SearchTopoContainedEW}{21.1\% [20.8, 21.4]}
\newcommand{\SearchTopoContainedUW}{19.5\% [19.2, 19.8]}
\newcommand{\SearchTopoOpenerEW}{15.4\% [15.3, 15.5]}
\newcommand{\SearchTopoOpenerUW}{17.9\% [17.6, 18.1]}
\newcommand{\SearchTopoCloserEW}{9.8\% [9.7, 9.9]}
\newcommand{\SearchTopoCloserUW}{9.2\% [9.0, 9.3]}
\newcommand{\SearchTopoBridgeEW}{53.7\% [53.3, 54.0]}
\newcommand{\SearchTopoBridgeUW}{53.5\% [53.1, 53.8]}
\newcommand{\SearchCompContentAnywhereEW}{78.9\% [78.6, 79.2]}
\newcommand{\SearchCompContentAnywhereUW}{80.5\% [80.2, 80.8]}
\newcommand{\SearchCompContentBeforeEW}{44.2\% [43.9, 44.5]}
\newcommand{\SearchCompContentBeforeUW}{46.7\% [46.4, 47.1]}
\newcommand{\SearchCompContentAfterEW}{54.2\% [53.9, 54.5]}
\newcommand{\SearchCompContentAfterUW}{60.2\% [59.9, 60.6]}
\newcommand{\WithinUserContainedDiff}{+13.0 [12.5, 13.6]}
\newcommand{\WithinUserAIDirection}{+7.0 [6.4, 7.7]}
\newcommand{\WithinUserSearchDirection}{-13.5 [-13.9, -13.2]}
\newcommand{\DirectionalRecomposition}{+20.6 [19.9, 21.3]}

\newcommand{\AllSearchContainedUW}{18.2\% [17.9, 18.4]}

\newcommand{\AllSearchBridgeUW}{55.6\% [55.2, 55.9]}
\newcommand{\CompAIDurationMin}{22}

\newcommand{\CompAISwitchesCI}{5.2 [5.0, 5.4]}
\newcommand{\CompAIAssistantShare}{59.7\%}
\newcommand{\CompAISearchShare}{8.8\%}
\newcommand{\CompAIContentShare}{31.5\%}
\newcommand{\CompSearchDurationMin}{9}

\newcommand{\CompSearchSwitchesCI}{4.0 [3.9, 4.1]}
\newcommand{\CompSearchSearchShare}{48.6\%}
\newcommand{\CompSearchContentShare}{51.4\%}
\newcommand{\BridgeBothSidesEW}{74.0\% [73.3, 74.8]}

\newcommand{\BridgeBetweenOnlyEW}{26.0\% [25.3, 26.6]}

\newcommand{\BridgeCoincidentOnlyEW}{0.0\% [0.0, 0.0]}

\newcommand{\NextWebFiveMinCI}{21.8\% [21.2, 22.3]}

\newcommand{\NextWebFifteenMinCI}{32.3\% [31.6, 33.0]}

\newcommand{\NextWebThirtyMinCI}{39.7\% [39.1, 40.5]}

\newcommand{\NextWebOneHourCI}{50.5\% [49.7, 51.4]}
\newcommand{\NextWebOneDay}{90.5\%}

\newcommand{\NextSearchThirtyMinCI}{19.1\% [18.7, 19.6]}

\newcommand{\NextSearchOneHourCI}{26.8\% [26.2, 27.5]}

\newcommand{\ContainedNextWebOneHourCI}{12.0\% [11.6, 12.4]}

\newcommand{\ContainedNextWebSixHourCI}{49.5\% [48.3, 50.6]}
\newcommand{\GapFifteenContained}{46.9\%}
\newcommand{\GapFifteenOpener}{9.1\%}
\newcommand{\GapFifteenCloser}{14.7\%}
\newcommand{\GapFifteenBridge}{29.4\%}
\newcommand{\GapThirtyContained}{37.8\%}
\newcommand{\GapThirtyOpener}{9.6\%}
\newcommand{\GapThirtyCloser}{15.3\%}
\newcommand{\GapThirtyBridge}{37.3\%}
\newcommand{\GapSixtyContained}{29.2\%}
\newcommand{\GapSixtyOpener}{9.0\%}
\newcommand{\GapSixtyCloser}{14.9\%}
\newcommand{\GapSixtyBridge}{46.9\%}
\newcommand{\GapMaxCIHalfWidth}{1.0}
\newcommand{\SearchGapFifteenContained}{25.0\%}
\newcommand{\SearchGapSixtyContained}{16.4\%}
\newcommand{\SearchGapMaxCIHalfWidth}{0.4}
\newcommand{\CoverageOneContained}{37.7\%}
\newcommand{\CoverageOneOpener}{9.6\%}
\newcommand{\CoverageOneCloser}{15.3\%}
\newcommand{\CoverageOneBridge}{37.3\%}
\newcommand{\CoverageSevenContained}{35.3\%}
\newcommand{\CoverageSevenOpener}{9.9\%}
\newcommand{\CoverageSevenCloser}{16.0\%}
\newcommand{\CoverageSevenBridge}{38.9\%}
\newcommand{\CoverageFourteenContained}{33.2\%}
\newcommand{\CoverageFourteenOpener}{10.0\%}
\newcommand{\CoverageFourteenCloser}{16.5\%}
\newcommand{\CoverageFourteenBridge}{40.3\%}
\newcommand{\CoverageTwentyOneContained}{31.7\%}
\newcommand{\CoverageTwentyOneOpener}{10.3\%}
\newcommand{\CoverageTwentyOneCloser}{17.3\%}
\newcommand{\CoverageTwentyOneBridge}{40.7\%}
\newcommand{\CoverageTwentyEightContained}{26.7\%}
\newcommand{\CoverageTwentyEightOpener}{10.9\%}
\newcommand{\CoverageTwentyEightCloser}{18.9\%}
\newcommand{\CoverageTwentyEightBridge}{43.5\%}
\newcommand{\CoverageMaxCIHalfWidth}{1.2}
\newcommand{\MarketUSAIContained}{50.9\%}
\newcommand{\MarketUSSearchContained}{21.6\%}
\newcommand{\MarketUSAIDirection}{+1.9 [+1.1, +2.7]}
\newcommand{\MarketUSSearchDirection}{-18.3 [-18.9, -17.8]}
\newcommand{\MarketGBAIContained}{20.7\%}
\newcommand{\MarketGBSearchContained}{17.9\%}
\newcommand{\MarketGBAIDirection}{+10.3 [+9.5, +11.0]}
\newcommand{\MarketGBSearchDirection}{-9.8 [-10.2, -9.4]}
\newcommand{\AsstChatGPTContained}{27.3\%}
\newcommand{\AsstChatGPTDirection}{+9.1 [+8.4, +9.8]}
\newcommand{\AsstChatGPTSearchDirection}{-12.4 [-12.8, -12.0]}
\newcommand{\AsstGeminiContained}{48.8\%}
\newcommand{\AsstGeminiDirection}{+1.2 [+0.2, +2.2]}
\newcommand{\AsstGeminiSearchDirection}{-16.2 [-16.9, -15.5]}

\newcommand{\SearchDefCanonicalAnywhereEW}{42.1\%}
\newcommand{\SearchDefCanonicalAnywhereUW}{46.0\%}
\newcommand{\SearchDefCanonicalCompContained}{21.1\%}
\newcommand{\SearchDefCanonicalWithinDiff}{+13.0}
\newcommand{\SearchDefKeyPhraseAnywhereEW}{39.6\%}
\newcommand{\SearchDefKeyPhraseAnywhereUW}{43.3\%}
\newcommand{\SearchDefKeyPhraseCompContained}{19.9\%}
\newcommand{\SearchDefKeyPhraseWithinDiff}{+13.6}
\newcommand{\SearchDefGBDAnywhereEW}{42.0\%}
\newcommand{\SearchDefGBDAnywhereUW}{45.8\%}
\newcommand{\SearchDefGBDCompContained}{21.1\%}
\newcommand{\SearchDefGBDWithinDiff}{+13.0}
\newcommand{\SearchDefNoYahooRootAnywhereEW}{42.1\%}
\newcommand{\SearchDefNoYahooRootAnywhereUW}{45.9\%}
\newcommand{\SearchDefNoYahooRootCompContained}{21.1\%}
\newcommand{\SearchDefNoYahooRootWithinDiff}{+13.0}
\newcommand{\RegDomainAnywhereEW}{46.6\%}
\newcommand{\RegDomainAnywhereUW}{50.6\%}
\newcommand{\RegDomainNonSearchShare}{21.6\%}
\newcommand{\AltEngineMaxShare}{0.03\%}

\newcommand{\CrossFebAIContained}{34.1\% [33.5, 34.7]}

\newcommand{\CrossFebSearchContained}{19.5\% [19.2, 19.8]}

\newcommand{\CrossFebWithinContained}{+13.0 [12.5, 13.6]}
\newcommand{\CrossFebDirectional}{+20.6 [19.9, 21.3]}
\newcommand{\CrossMarAIContained}{35.3\% [34.8, 36.0]}
\newcommand{\CrossMarAIBridge}{36.2\% [35.6, 36.7]}
\newcommand{\CrossMarSearchContained}{20.4\% [20.1, 20.7]}
\newcommand{\CrossMarSearchBridge}{52.8\% [52.5, 53.2]}
\newcommand{\CrossMarWithinContained}{+13.6 [13.0, 14.1]}
\newcommand{\CrossMarDirectional}{+20.3 [19.6, 20.9]}

\newcommand{\TaskShoppingDiff}{+19.8 [+17.7, +22.3]}
\newcommand{\TaskNewsDiff}{+21.1 [+17.8, +23.9]}
\newcommand{\TaskCodingDiff}{+18.4 [+11.4, +25.7]}
\newcommand{\TaskReferenceDiff}{+26.8 [+22.8, +31.4]}
\newcommand{\TaskLeisureDiff}{+15.4 [+13.6, +17.2]}
\newcommand{\TaskOtherDiff}{+21.8 [+21.1, +22.6]}
\newcommand{\TaskPooledCategorizedDiff}{+18.3 [+17.1, +19.5]}
\newcommand{\TaskCoverageAI}{22.5\%}
\newcommand{\TaskCoverageSearch}{24.3\%}

\setcopyright{none}
\acmConference[Preprint]{Manuscript under review --- not yet peer reviewed}{2027}{}
\acmYear{2027}

\begin{document}

\title{The New Shape of Search: How Conversational AI Recomposes
Information Seeking}

\ifanon
\author{Anonymous Author(s)}
\else
\author{Michael Iannelli}
\affiliation{\institution{Scrunch AI}\city{New York}\country{USA}}
\email{michael@scrunchai.com}
\author{Alan Ai}
\affiliation{\institution{Scrunch AI}\city{New York}\country{USA}}
\email{alan.ai@scrunchai.com}
\renewcommand{\shortauthors}{Iannelli and Ai}
\fi

\begin{abstract}
The familiar search journey begins with a query and moves outward into documents, and
conversational AI is commonly imagined at its mouth: ask first, then click out. Linking
captured prompts and responses to the same panelists' observed searches and pageviews, and
reconstructing inactivity-defined cross-surface temporal sessions (standalone assistant
surfaces; search-embedded AI such as AI Overviews and AI Mode is out of scope, since it
co-occurs with the results page), we find the observed
journeys more often run the other way. Content usually follows search but more often
precedes assistant use. Within the same panelist, the paired difference-of-directions
between the two anchors is \DirectionalRecomposition{} percentage points; it persists
within every coarse destination-domain stratum we can observe (semantic task and task-stage
matching remain unresolved), and every headline result replicates in a
second, adjacent month. Search tends to anchor the front of the observed journey;
assistants sit deeper inside it.

Assistant sessions are also far more often self-contained. User-weighted,
\TopoContainedUW{} of assistant-containing sessions show no observed external web step
(AI-first \TopoOpenerUW{}, AI-last \TopoCloserUW{}, bridge/interleaved \TopoBridgeUW{}),
against \SearchTopoContainedUW{} contained for search-centered sessions of the same users,
a within-user contrast of \WithinUserContainedDiff{} percentage points. We call this
difference \emph{recomposition}: activity is distributed differently across dialogue,
search, and browsing, without implying that assistant use caused the difference.
Assistant-contained also does not mean resolved: timestamps alone cannot establish one
task, satisfaction, or completion. The result is a cross-surface topology of the emerging
search journey and a discipline for distinguishing observed containment from inferred
resolution.
\end{abstract}



\maketitle
\raggedbottom

\section{Introduction}
A person wants to understand a medical result, choose a car seat, or follow a breaking
story. In the model that has organized information-retrieval research for decades, they
begin in uncertainty, issue a query, scan a ranked list, reformulate, gather across sources,
and synthesize, an iterative multi-step \emph{episode} instead of a single
lookup~\cite{kuhlthau1991,marchionini2006exploratory,belkin1982ask}. Conversational AI is now
inserted into that episode. A common framing of what it does is the \emph{answer engine}:
a prompt goes in, a synthesized answer comes out, the episode ends.

That framing has an appealing empirical signature. If a clickstream shows no onward search
or pageview after an assistant response, it is natural to call the episode finished. But
this reading depends on two decisions that do more work than they appear to. First, it
treats prompts and responses as non-actions, so a provider session with several exchanges
looks the same as one response. Second, it treats a lack of outward web activity as evidence
about the information need rather than as evidence only about the observed surface. Counting
conversational events is necessary; calling them one task or a successful resolution requires
additional semantic evidence.

We take a more structural view. The familiar observable shape of search begins with a
query and moves outward: results, documents, reformulations, and synthesis. Conversational
AI introduces another place where that work can occur. People can begin with the assistant,
arrive there after encountering the web, remain inside dialogue, or move repeatedly between
surfaces. The consequential question is therefore not simply whether an assistant replaces
a query. It is where the assistant sits inside the broader observed journey.

Two surfaces carry this AI. \emph{Search-embedded} AI---Google's AI Overviews and AI
Mode---renders a synthesized answer inside the results page, with no session separable from the
query it sits in. A \emph{standalone} conversational assistant---ChatGPT, Claude, Perplexity, or
Gemini's own surface---is a destination the user navigates to, generating a session with its own
surrounding web context. The before-and-after web context we measure is therefore defined only for
the standalone case. We study the standalone surfaces; the in-SERP surface has its own measurable
behavior, such as reduced onward clicking when a synthesized answer
appears~\cite{pew2025aioverviews}, that a within-SERP instrument rather than ours is suited to
observe, and we treat it as a companion object requiring a different measurement design.

We call a difference in that distribution \emph{recomposition}: information-seeking
activity is arranged differently across dialogue, search, and browsing. This is a
descriptive claim, not a causal one. The comparison cannot separate what assistants change
from the kinds of needs people choose to bring to them. Using an opt-in panel that links
assistant activity to the same panelists' observed search and browsing, we ask a question
that classic web logs cannot answer and assistant-only corpora cannot either:
\emph{where does external web activity fall around assistant use, and how does that differ
from where it falls around conventional search?} Our thesis is:

\begin{quote}
Search tends to open the observed journey toward content, while assistant use more often
follows prior web activity. Measured with one construction on both anchors in the same
users, the two shapes differ in direction and in containment, and no single ``answer then
exit'' pattern describes the observed system.
\end{quote}

\begin{figure*}[t]
\centering
\includegraphics[width=0.92\textwidth]{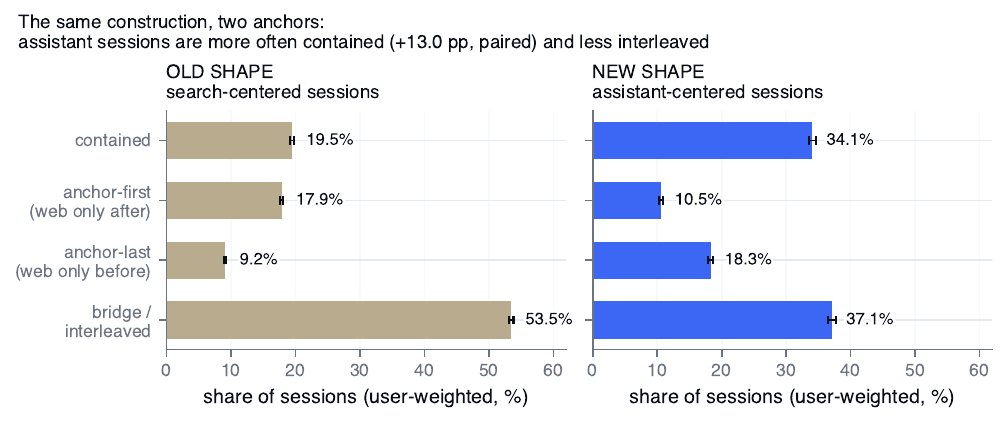}
\caption{Old shape versus new shape: the same four-position temporal topology constructed
around conventional search (left; non-assistant sessions of the same users) and around
conversational assistants (right). Matched definitions, 30-minute gap, user-weighted shares
with 95\% user-clustered bootstrap intervals. The comparison is descriptive: each anchor
selects its own sessions, and position does not establish task identity or causation.}
\Description{Two mirrored horizontal bar charts sharing one axis. The left panel shows
search-centered sessions and the right panel assistant-centered sessions, each with four
bars: contained, anchor-first, anchor-last, and bridge or interleaved. Exact values appear
as labels on each bar and come from the tracked result manifest.}
\label{fig:shapes}
\end{figure*}

We make this concrete twice with one construction. First, a topology of AI-containing
temporal sessions built from the
position of assistant events relative to the web (Figure~\ref{fig:topology}). We use
observable labels---contained, AI-first, AI-last, and bridge/interleaved---because the trace
tells us where events sit, not what they caused or whether they share one task. Second, the
\emph{identical} construction centered on conventional search in the same users' non-assistant
sessions (\S\ref{sec:comparator}): the old shape, measured with the new instrument, so the
two shapes can be compared like for like (Figure~\ref{fig:shapes}).

\paragraph{Contributions.}
(1) A method and unit: \textbf{cross-surface session reconstruction} that counts prompts and
responses as first-class events alongside searches and pageviews, from a same-user panel,
together with the direct positional marginals and the gap, coverage-density, and
time-to-next-event diagnostics that expose which conclusions depend on session construction
(\S\ref{sec:data}, \S\ref{sec:topology}, Appendix~\ref{sec:permeability}).
(2) A \textbf{full-session topology} of AI-containing sessions, contained, AI-first,
AI-last, and bridge/interleaved, reported both session-weighted
and user-weighted with user-clustered intervals, stable in its AI-first and
AI-last shares while the contained/bridge boundary moves with the segmentation gap; inside
containment, a construct-validity decomposition shows that multiple captured responses do
not by themselves establish one dialogue or successful resolution
(\S\ref{sec:topology}, \S\ref{sec:contained}).
(3) An \textbf{old-shape comparator}: the same temporal topology centered on conventional
search among assistant adopters, with within-user contrasts of containment and direction;
the paired reversal---content usually follows search but more often precedes assistant
use---is the paper's central evidence of recomposition
(\S\ref{sec:comparator}).

\section{Related Work}
\paragraph{Information seeking is an episode, not a query.} Classic models cast search as an
affective and cognitive process from uncertainty toward focus instead of mechanical
retrieval~\cite{kuhlthau1991,kuhlthau2004}, often beginning from an anomalous state of
knowledge~\cite{belkin1982ask} or a sense-making gap~\cite{dervin1998sensemaking,wilson1999models}.
Exploratory search rejects the single-query model~\cite{marchionini2006exploratory}, strategies
vary within one episode~\cite{belkin1995cases}, the session rather than the query is the natural
unit of analysis and evaluation~\cite{jarvelin2008session}, queries are reformulated as the user
learns~\cite{huang2009reformulation}, search itself is a form of
learning~\cite{rieh2016learning,vakkari2016learning}, and foraging accounts describe how seekers
trade cost against value in deciding where to look~\cite{pirolli1999foraging}. Conversational
search formalizes the multi-turn answer-then-continue loop and treats follow-up turns as
meaningful actions~\cite{radlinski2017framework}. We treat the task episode as the conceptual unit
but the inactivity-defined temporal session as the observed proxy; this distinction is central to
the paper's claim discipline. Inactivity-defined sessions are a long-standing measurement
compromise whose thresholds are heuristic rather than behaviorally
derived~\cite{catledge1995characterizing}; our gap and coverage sensitivities quantify exactly how much of
the cross-surface topology depends on that compromise.

\paragraph{Taxonomies of search, by intent and by surface.} Web search has long been classified by
\emph{intent}---Broder's navigational, informational, and transactional
split~\cite{broder2002taxonomy}, recently revised for the generative era into knowledge-,
guidance-, and output-seeking intents that deliberately span both search engines and AI
chatbots~\cite{lichtenegger2026taxonomy}. Conversational information seeking is itself defined by
multi-turn dialogue rather than by where a surface sits~\cite{zamani2023cis}, and the ``answer
engine'' framing names the synthesized-answer output without distinguishing its
locus~\cite{shah2022situating}. Our cut is complementary and measurement-driven: a
\emph{search-embedded} surface, whose answer has no session separable from the query it sits in,
versus a \emph{standalone} surface that generates a session with surrounding web context. The
before/after topology we measure is defined only for the latter, which is why our instrument
studies the standalone surface and leaves the embedded one to a within-SERP design.

\paragraph{Does AI displace search, and where does behavior go?} Information seeking is among the
most common conversational-AI uses~\cite{chatterji2025chatgpt}, and a large same-user before/after
study finds no drop in search usage after assistant adoption~\cite{semrush2025expansion}. Aggregate
trend claims are hard to identify because adoption coincides with activity
bursts~\cite{barabasi2005bursts} that inflate volume comparisons. A methods companion on this panel
makes the obstacle concrete: assistant use is itself timed by the user, so known-null timestamps
drawn at comparably active moments reproduce much of the apparent post-event search
lift~\cite{iannelli2026burst}. We therefore avoid a volume verdict and study episode composition
instead. A parallel concern is where AI concentrates the sources
people encounter: audits report that AI answers draw on fewer, more central
sources~\cite{aral2026aisearch} and produce engine-specific answer
bubbles~\cite{huang2026answerbubbles}. These measure system outputs; we measure where the same users
actually go, at episode scope.

\paragraph{AI behavior in the wild.} Conversation corpora establish what people
ask~\cite{zhao2024wildchat,yu2026searchable} but lack the same-user link to subsequent web behavior;
answer-engine audits study outputs without users. We supply the behavioral middle at the same-user,
cross-surface event level. Prior work on cognitive offloading to external
tools~\cite{sparrow2011google} and on lost-in-the-middle attention in long
contexts~\cite{liu2024lost} motivates later content validation rather than functional labels here.
\ifanon
Prior public work on a comparable cross-surface panel develops the retail-demand
channel~\cite{iannelli2026fptp}, and a public methods preprint develops the event-time
confounding that keeps this paper off volume claims~\cite{iannelli2026burst}; here the object is
the seeking episode itself.
\else
A public companion preprint on the same panel, accepted at a KDD~2026 workshop, develops the
retail-demand channel~\cite{iannelli2026fptp}, and a second public preprint develops the event-time
confounding that keeps this paper off volume claims~\cite{iannelli2026burst}; here the object is
the seeking episode itself. Both companions are public, so nothing here rests on work a reader
cannot check.
\fi

\section{Data and Cross-Surface Session Reconstruction}
\label{sec:data}
\paragraph{Panel and measurement instrument.} An opt-in cross-surface research panel, whose members
consent to have their device activity metered for research, covers February 2026 in the United States
and Great Britain. The provider's metering software records two browser-level event streams for the
same user. The \emph{pageview stream} records user-facing page visits with a timestamp, the host-level
URL domain, the referrer domain, a tab identifier, and active dwell time; it reflects navigations the
browser rendered for the user, not background network requests. The \emph{assistant stream} records
conversation events on metered assistant web surfaces: each captured prompt and each captured response
carries a timestamp, its role, a provider label, a provider session identifier, and content length.
Assistant use through unmetered channels (for example native mobile apps or API access) is not
observed, which is one reason we treat absence of events as absence of \emph{observation}, never as
absence of activity. The \emph{standalone conversational assistants} we study---the \emph{elected} set: ChatGPT,
Claude, Perplexity, and Gemini's standalone surface---are surfaces the user
deliberately navigates to. Google's search-embedded AI (AI Overviews and AI Mode) is out of scope
for the structural reason set out in the introduction: a search-embedded answer has no session
separable from the query it sits in, so it co-occurs with the search
event by construction, cannot be separated cleanly in our capture, and its shape requires a
different, within-SERP measurement design.

\paragraph{Population and denominator flow.} The target frame contains every panelist in the window
with at least one \emph{complete} elected-assistant provider session, meaning a provider session with
at least one non-empty captured prompt and at least one non-empty captured response. From that frame,
the primary cross-surface analysis requires at least one observed pageview in the month; this retains
nearly all eligible assistant-session users. Event presence is not guaranteed continuous
instrumentation, so we additionally report how the headline shares move as the required number of
pageview-active days rises (Table~\ref{tab:coverage}). To protect session construction at the month
boundaries, events are pulled with a sixty-minute buffer on each side of February (one maximum
inactivity gap), the stream is sessionized on the buffered pull, and a session enters the analysis
if and only if its first event falls inside February; retained sessions keep their out-of-month
tails instead of being truncated. The next-event analysis additionally censors at the window end
under a strict risk-set rule (Appendix~\ref{sec:permeability}). In this window the elected-assistant
events are ChatGPT and Gemini with a small Perplexity share; only providers with eligible observed
sessions in this window contribute to provider-specific estimates, and the extraction fails loudly
if any provider outside the elected set ever appears.

\paragraph{Session reconstruction with assistant events as steps.} A dedicated paper-owned extract
hashes user and provider-session identifiers before transfer and retains timestamps, prompt/response
roles, sequence, assistant label, and content length---never prompt or response text. It retains coarse
pageview and referrer domains but no URL path, URL query, title, search phrase, or page content. We
interleave each user's prompts, responses, searches, and content pageviews into one time-ordered stream
and segment it at a 30-minute inactivity gap; the gap sensitivities in Table~\ref{tab:gap} are
recomputed from the raw event stream, not by merging or splitting existing sessions. We call the
resulting object a \emph{cross-surface
temporal session}; it is a proxy for, not proof of, a coherent task episode. A provider session can
span an inactivity boundary, so one temporal session may contain only part of an eligible provider
session's events. Pageviews to assistant
surfaces are dropped from the web stream so they are not double-counted against prompt/response events. A
web step is a search (a pageview whose full host is a canonical search-engine
surface\footnote{The canonical search-host list: google.com, www.google.com, m.google.com,
google.co.uk, www.google.co.uk, images.google.com, images.google.co.uk, bing.com, www.bing.com,
duckduckgo.com, www.duckduckgo.com, yahoo.com, www.yahoo.com, search.yahoo.com. Alternative engines
absent from the list (Ecosia, Brave, Startpage, Yandex, AOL) each account for at most \AltEngineMaxShare{} of
observed web events in this window. Matching by \emph{registrable domain} instead would classify
mail, document, calendar, translation, and app-store hosts served from google.com and yahoo.com as
search---\RegDomainNonSearchShare{} of the events flagged by the registrable-domain rule---and would inflate search-anywhere from
\SearchAnywhereEW{} to \RegDomainAnywhereEW{} session-weighted and from \SearchAnywhereUW{} to \RegDomainAnywhereUW{} user-weighted.})
or a content
pageview (any other non-chat pageview). Throughout, ``web'' and ``external'' cover both kinds of
step: search pageviews plus all other non-chat pageviews. Timestamps are capture times recorded by the metering software at sub-second resolution; for
responses we do not observe generation start versus completion separately, so the assistant
span reflects when events were captured on the surface, and positions within a few seconds of
a span boundary inherit that imprecision. An
external event exactly coincident with an assistant event does not identify an ordering, so it makes
the session bridge/interleaved rather than fabricating a before/after position (coincidences are
rare: well under 0.1\% of sessions). Remaining same-timestamp orderings affect only display
order, never classification, which depends on inequalities alone. A session is \emph{AI-containing} if it holds at least one
prompt or response event from an eligible provider session.

\paragraph{The topology.} For each AI-containing temporal session we locate the assistant span (first
to last prompt/response event) and ask where external web steps fall relative to it: before the span,
after it, or between assistant events. This yields four mutually exclusive temporal classes
(Table~\ref{tab:topology}):
\emph{assistant-contained} (no external step anywhere),
\emph{AI-first} (external only after the span),
\emph{AI-last} (external only before the span), and
\emph{bridge/interleaved} (external on both sides, between assistant events, or coincident with an
assistant event). Between-only sessions belong to bridge but have nothing before the first assistant
event, so before/after marginals are estimated directly rather than reconstructed by adding classes.

\paragraph{Four discretionary choices.} Four choices are genuinely discretionary, and we fix
defensible defaults instead of hiding them. (i) The 30-minute inactivity gap defines temporal-session
boundaries; it is the conventional web-sessionization default and the middle of the sensitivity range
we report at 15, 30, and 60 minutes (Table~\ref{tab:gap}). (ii) Provider
\texttt{session\_id} defines an assistant session; it is a recorded grouping key, and the
cross-surface inactivity rule may combine several provider sessions or split one. (iii) Search is
defined by full host against the canonical search-host list in the footnote above; the registrable-
domain alternative and its measured effect are reported there. (iv) The primary population requires
one pageview-active day, the least restrictive threshold of the ladder in Table~\ref{tab:coverage}; we
report thresholds through
all 28 days. Workbench-versus-seeking, topical continuity, satisfaction, and per-task splits require a
separately governed content-validation pass and are not inferred here.

\paragraph{The search label is a construct choice.} Because both the AI-side search
marginals and the entire comparator rest on what counts as ``search,'' we recompute the
load-bearing quantities under four defensible rules using only privacy-safe retained fields
(Table~\ref{tab:searchdef}). Requiring a captured key phrase is conservative (phrase capture
can fail on genuine searches) and dropping the Yahoo portal roots isolates portal-homepage
traffic. Every variant stays within a narrow band of the primary rule, none approaches the
rejected registrable-domain rule's inflated values (\RegDomainAnywhereEW{} / \RegDomainAnywhereUW{} search-anywhere), and
the comparator's containment level and the within-user contrast move by well under the
contrast itself.

\begin{table}[t]
\caption{Search-definition construct sensitivity (30-minute gap; point estimates).
Search-anywhere is within AI-containing sessions (session-/user-weighted); the last two
columns are the search-centered comparator's contained share (session-weighted) and the
within-user contained contrast (AI minus search, percentage points).}
\label{tab:searchdef}
\small\setlength{\tabcolsep}{3pt}
\begin{tabular}{lrrrr}
\toprule
definition & \multicolumn{2}{c}{search anywhere} & comp.\ cont. & within-user \\
\midrule
canonical (primary) & \SearchDefCanonicalAnywhereEW & \SearchDefCanonicalAnywhereUW & \SearchDefCanonicalCompContained & \SearchDefCanonicalWithinDiff \\
+ key phrase & \SearchDefKeyPhraseAnywhereEW & \SearchDefKeyPhraseAnywhereUW & \SearchDefKeyPhraseCompContained & \SearchDefKeyPhraseWithinDiff \\
Google/Bing/DDG & \SearchDefGBDAnywhereEW & \SearchDefGBDAnywhereUW & \SearchDefGBDCompContained & \SearchDefGBDWithinDiff \\
minus Yahoo roots & \SearchDefNoYahooRootAnywhereEW & \SearchDefNoYahooRootAnywhereUW & \SearchDefNoYahooRootCompContained & \SearchDefNoYahooRootWithinDiff \\
\bottomrule
\end{tabular}
\end{table}

\paragraph{Estimands and uncertainty.} Every headline share is reported under two
weightings.
\emph{Session-weighted} is the mean over AI-containing temporal sessions.
\emph{User-weighted} is the equal-weight mean of per-user shares: each user's sessions are
first averaged within user, then users are averaged with equal weight regardless of how many
sessions they contribute. Uncertainty comes from a nonparametric cluster bootstrap that
resamples \emph{panelists} (never individual sessions or events) with replacement, 500
replicates, percentile 95\% intervals, with the full weighting recomputed inside each
replicate. The topology, positional-marginal, contained-decomposition, and bridge-composition
estimates carry these intervals in the text and tables. The gap and coverage tables display
point estimates for readability; their user-clustered intervals are computed identically, are
recorded in the tracked result manifest, and have half-widths of at most
\GapMaxCIHalfWidth{} and \CoverageMaxCIHalfWidth{} percentage points respectively. The
next-event curves carry user-clustered bands (Figure~\ref{fig:continuation}); the values
quoted in Appendix~\ref{sec:permeability} include their intervals.

\paragraph{Disclosure and reproducibility.} We study standalone conversational assistants linked to the same
users' search and browsing in the United States and Great Britain. We do not report aggregate panel size,
per-analysis user counts, or per-cell sample sizes; these are commercially sensitive. No raw
conversations, searches, or individual URLs are released. A tracked result manifest records every
reported share and interval; LaTeX values and figures are generated from that manifest, over the
unsampled eligible frame.

\paragraph{On causality.} We make no causal-volume claim. Every quantity is descriptive of temporal
session structure. Selection into assistant use and the semantic relationship among co-timed events
are not identified away.

\section{The Topology of AI-Containing Temporal Sessions}
\label{sec:topology}
\begin{figure*}[t]
\centering
\includegraphics[width=0.92\textwidth]{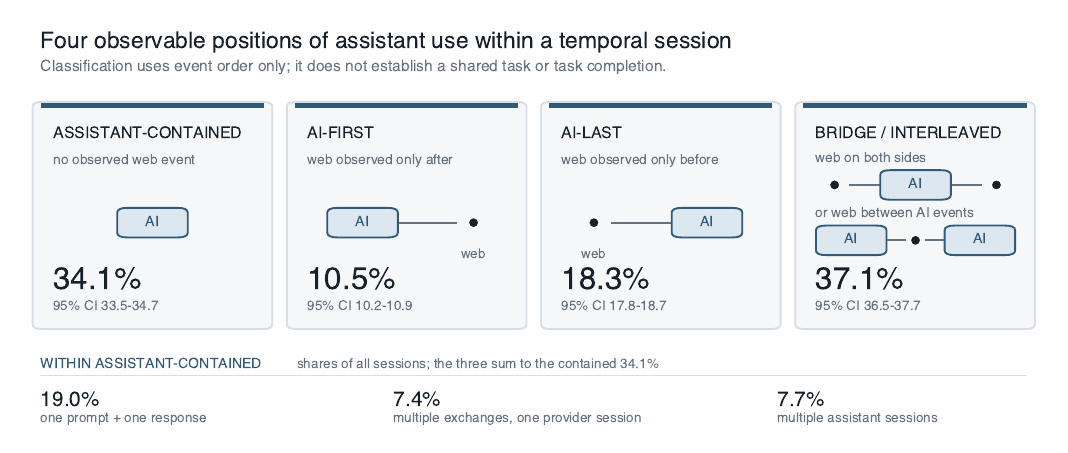}
\caption{Cross-surface topology of AI-containing temporal sessions. A session is classified by where
external web activity falls relative to the assistant-event span: no observed external web step within
the constructed session (assistant-contained), only
after (AI-first), only before (AI-last), or on both sides / between assistant events
(bridge/interleaved). Contained sessions are subdivided using observed prompt, response, and
provider-session identifiers, not inferred resolution. Cards show user-weighted shares with 95\%
user-clustered bootstrap intervals; 30-minute inactivity gap.}
\Description{A four-card taxonomy of observable event positions. Assistant-contained sessions,
with no observed web event, account for 34.1 percent; AI-first sessions, with web only after, 10.5
percent; AI-last sessions, with web only before, 18.3 percent; and bridge or interleaved sessions,
with web on both sides of the assistant span or between assistant events, 37.1 percent. Sequence
glyphs depict each event order. A secondary row divides the contained share into one prompt plus
one response, 19.0 percent; multiple exchanges in one provider session, 7.4 percent; and multiple
assistant sessions, 7.7 percent. The figure states that event order does not establish a shared
task or task completion.}
\label{fig:topology}
\end{figure*}

The cross-surface topology is the paper's central object (Figure~\ref{fig:topology},
Table~\ref{tab:topology}). Assistant-contained accounts for \TopoContainedEW{} of sessions
session-weighted and \TopoContainedUW{} user-weighted. Bridge/interleaved is similarly common
(\TopoBridgeEW{} and \TopoBridgeUW{}), followed by AI-last (\TopoCloserEW{} and
\TopoCloserUW{}) and AI-first (\TopoOpenerEW{} and \TopoOpenerUW{}). Even the largest
class holds barely more than a third of sessions.

\begin{table}[t]
\caption{Temporal-session topology shares (30-minute gap), session-weighted and user-weighted, with
95\% user-clustered bootstrap intervals. Contained subclasses are fully observable event/session
patterns reported as shares of \emph{all} AI-containing sessions (they sum to the contained share):
one prompt + one response means exactly those two captured events in one provider session; multiple
exchanges means any other event pattern within one provider session; multiple assistant sessions
means events from more than one provider session. Displayed classes may not sum exactly to 100\%
because of rounding.}
\label{tab:topology}
\small\setlength{\tabcolsep}{3pt}
\begin{tabular}{lcc}
\toprule
position & session-weighted & user-weighted \\
\midrule
assistant-contained                  & \TopoContainedEW & \TopoContainedUW \\
\quad one prompt + one response      & \TopoContainedSingleExchangeEW & \TopoContainedSingleExchangeUW \\
\quad multiple exchanges, one session & \TopoContainedMultiExchangeEW & \TopoContainedMultiExchangeUW \\
\quad multiple assistant sessions    & \TopoContainedMultipleSessionsEW & \TopoContainedMultipleSessionsUW \\
AI-first (web only after)             & \TopoOpenerEW & \TopoOpenerUW \\
AI-last (web only before)             & \TopoCloserEW & \TopoCloserUW \\
bridge / interleaved                  & \TopoBridgeEW & \TopoBridgeUW \\
\bottomrule
\end{tabular}
\end{table}

\paragraph{Position must be estimated directly.} Bridge contains between-only sessions, so adding
AI-last and bridge would overstate the share with web activity before the first assistant event. The
direct marginal is \WebBeforeEW{} session-weighted and \WebBeforeUW{} user-weighted. The
four-to-five-point gap between the estimands is itself informative: highly active users contribute more
sessions, while an
equal-user estimand gives more weight to lighter users. Search appears somewhere in
\SearchAnywhereEW{} of sessions session-weighted and \SearchAnywhereUW{} user-weighted; specifically,
it appears before the first assistant event in \SearchBeforeEW{} / \SearchBeforeUW{}, after the last
in \SearchAfterEW{} / \SearchAfterUW{}, and between assistant events in
\SearchBetweenEW{} / \SearchBetweenUW{}. The positional marginals are not mutually exclusive---one
session can contribute to before and after at once---so they do not sum to the anywhere share. These
are temporal positions, not evidence that the assistant
synthesized prior pages or caused later search.

\paragraph{Inside bridge/interleaved.} The largest class is not homogeneous. Session-weighted within
bridge sessions, \BridgeBothSidesEW{} have external web activity on both sides of the assistant span,
and \BridgeBetweenOnlyEW{} interleave web activity strictly between assistant events without
enclosing the span; timestamp coincidences alone account for \BridgeCoincidentOnlyEW{}. Both-sided
sessions are consistent with the web-to-AI-to-web pattern the bridge label suggests, while
between-only sessions are assistant activity punctuated by browsing; the topology keeps them under
one headline class because both defeat any one-sided reading.

\paragraph{What holds, and what moves with the gap.} Because a single inactivity threshold is a
modeling choice, we recompute shares at 15, 30, and 60 minutes (Table~\ref{tab:gap}). AI-first remains
9--10\% and AI-last remains near 15\%. What moves is the contained/bridge boundary: contained falls
from \GapFifteenContained{} to \GapSixtyContained{}, while bridge rises from
\GapFifteenBridge{} to \GapSixtyBridge{}. The stable result is the presence of several positions and
the relative stability of the one-sided classes, not one universal contained share.

\begin{table}[t]
\caption{Gap sensitivity of the four topology positions (session-weighted shares), recomputed from the
raw event stream at each gap. AI-first and AI-last
are stable; the contained/bridge boundary moves with the inactivity gap. User-clustered 95\% interval
half-widths are at most \GapMaxCIHalfWidth{} points for every cell (intervals in the tracked
manifest).}
\label{tab:gap}
\small\setlength{\tabcolsep}{6pt}
\begin{tabular}{lrrrr}
\toprule
inactivity gap & contained & AI-first & AI-last & bridge \\
\midrule
15 min & \GapFifteenContained & \GapFifteenOpener & \GapFifteenCloser & \GapFifteenBridge \\
30 min & \GapThirtyContained & \GapThirtyOpener & \GapThirtyCloser & \GapThirtyBridge \\
60 min & \GapSixtyContained & \GapSixtyOpener & \GapSixtyCloser & \GapSixtyBridge \\
\bottomrule
\end{tabular}
\end{table}

\paragraph{Measurement density also moves containment.} Requiring more pageview-active days (days
with a non-chat pageview) makes external events more likely to be observed. Contained falls from \CoverageOneContained{} with at least
one active day to \CoverageTwentyEightContained{} among users active on all 28 days; bridge rises from
\CoverageOneBridge{} to \CoverageTwentyEightBridge{} (Table~\ref{tab:coverage}). This is not a
preferred-sample ladder---heavier
browsers are a selected population that plausibly differs in tasks and habits, not a
denoised version of the frame---but it shows why containment must always be reported with an
instrumentation/coverage rule.

\begin{table}[t]
\caption{Coverage sensitivity: topology shares (session-weighted, 30-minute gap) as the required
number of pageview-active days rises. Rows are selected subpopulations, not corrections of one
another; the one-day row differs from the headline frame by a hair because active days count
non-chat pageviews while frame membership counts any pageview. User-clustered 95\% interval
half-widths are at most \CoverageMaxCIHalfWidth{} points for
every cell (intervals in the tracked manifest).}
\label{tab:coverage}
\small\setlength{\tabcolsep}{5pt}
\begin{tabular}{rrrrr}
\toprule
min.\ active days & contained & AI-first & AI-last & bridge \\
\midrule
1  & \CoverageOneContained & \CoverageOneOpener & \CoverageOneCloser & \CoverageOneBridge \\
7  & \CoverageSevenContained & \CoverageSevenOpener & \CoverageSevenCloser & \CoverageSevenBridge \\
14 & \CoverageFourteenContained & \CoverageFourteenOpener & \CoverageFourteenCloser & \CoverageFourteenBridge \\
21 & \CoverageTwentyOneContained & \CoverageTwentyOneOpener & \CoverageTwentyOneCloser & \CoverageTwentyOneBridge \\
28 & \CoverageTwentyEightContained & \CoverageTwentyEightOpener & \CoverageTwentyEightCloser & \CoverageTwentyEightBridge \\
\bottomrule
\end{tabular}
\end{table}

\section{Inside Assistant-Contained Temporal Sessions}
\label{sec:contained}
Assistant-contained means only that no non-assistant web event falls inside the constructed temporal
session. Session-weighted within contained sessions, \ContainedMultipleResponse{} contain multiple
captured responses, but \ContainedMultipleAssistantSession{} combine multiple provider sessions.
(On that same within-contained base, single-exchange sessions are
\ContainedSingleExchangeWithinContained{}, so the two leave a residual well under half a point:
contained sessions with several captured events but at most one response, for example a provider
session split by the inactivity rule.
The \texttt{Topo*} subclass shares quoted elsewhere are shares of \emph{all} sessions, not of
contained ones, and are not addable to this figure.) User-weighted, the contained
share decomposes into \TopoContainedSingleExchangeUW{} with one prompt and one response,
\TopoContainedMultiExchangeUW{} with multiple exchanges in one provider session, and
\TopoContainedMultipleSessionsUW{} with multiple provider sessions.

This decomposition undercuts the tempting ``multi-response equals internalized search'' inference as
a general reading of the trace, although the data do not test an internalization construct directly.
Multiple responses establish more assistant activity, but multiple provider sessions may reflect
distinct tasks, revisitation, or parallel use. Even a single provider session does not reveal
satisfaction or task completion. Assistant-contained is therefore a measurement category, not a
behavioral outcome. A content-linked validation pass could later estimate seeking, workbench use,
same-task continuity, and satisfaction, but those labels are not assigned in the present taxonomy.

\section{The Old Shape: A Search-Centered Comparator}
\label{sec:comparator}
The topology so far shows where the web sits around assistants. On its own it cannot say
whether that shape is \emph{new}: perhaps any anchoring activity---search included---produces
the same positional mix. We therefore apply a \emph{matched} construction to the old anchor. In
the same users and the same month, we take every temporal session that contains at least one
search event and no assistant event, define the span from first to last search event, and
classify content pageviews against it: \emph{search-contained} (no content pageview anywhere),
\emph{search-first} (content only after the span), \emph{search-last} (content only before),
and \emph{bridge/interleaved} (both sides, between searches, or coincident)
(Figure~\ref{fig:shapes}). One event-ontology difference is unavoidable and we handle it
explicitly: around assistants the non-anchor events are search \emph{plus} content pageviews,
whereas around search the non-anchor events are content pageviews alone (search \emph{is} the
anchor). We therefore state the headline directional contrast content-only on both sides, so
the anchors are compared like for like, and report a sensitivity that instead keeps every
search-containing session (including those that also contain assistant events) below. The
comparator's sessions are selected by their own anchor---this
is a descriptive contrast between how two anchors organize their surroundings in the same
people, not a matched-task or historical comparison; the selection caveats are developed at
the end of this section.

\paragraph{Three contributor sets.} Three contributor sets are in play and we keep them
distinct. The assistant-centered estimates are computed over all AI-containing sessions of the
primary frame; their user-weighted form averages over every frame user, all of whom contribute
at least one such session by construction. The search-centered estimates are computed over the
qualifying non-assistant search sessions of the \emph{same} frame; their user-weighted form
averages over the frame users who contribute at least one such session (users with no
qualifying search-only session simply do not enter, and are not imputed). The within-user
contrast is restricted to the intersection: panelists contributing at least one session of
each type in the month. The comparator thus describes assistant adopters' own non-assistant
search sessions, never a pre-AI population or a historical baseline.

\paragraph{What differs.} Search-centered sessions are contained in \SearchTopoContainedEW{}
of cases session-weighted and \SearchTopoContainedUW{} user-weighted, against
\TopoContainedEW{} and \TopoContainedUW{} for assistant-centered sessions. The one-sided
classes are search-first \SearchTopoOpenerEW{} / \SearchTopoOpenerUW{} and search-last
\SearchTopoCloserEW{} / \SearchTopoCloserUW{}; bridge/interleaved is
\SearchTopoBridgeEW{} / \SearchTopoBridgeUW{}. Content appears somewhere in
\SearchCompContentAnywhereEW{} / \SearchCompContentAnywhereUW{} of search-centered sessions,
before the first search in \SearchCompContentBeforeEW{} / \SearchCompContentBeforeUW{} and
after the last in \SearchCompContentAfterEW{} / \SearchCompContentAfterUW{}. The
contained/bridge boundary moves with the inactivity gap here too (contained
\SearchGapFifteenContained{} at 15 minutes to \SearchGapSixtyContained{} at 60; interval
half-widths at most \SearchGapMaxCIHalfWidth{} points, intervals in the tracked manifest), so
the comparison holds the gap fixed rather than leaning on any one threshold.

\paragraph{The mirror.} The clearest structural difference is directional, and we state it
like-for-like: content pageviews on both sides. Around search, content mass sits
\emph{after} the anchor: content follows the last search in \SearchCompContentAfterUW{} of
sessions user-weighted against \SearchCompContentBeforeUW{} before the first. Around
assistants the asymmetry flips: content \emph{precedes} the first assistant event in
\ContentBeforeUW{} of sessions against \ContentAfterUW{} after the last. The same reversal
holds session-weighted (\SearchCompContentAfterEW{} versus \SearchCompContentBeforeEW{}
around search; \ContentBeforeEW{} versus \ContentAfterEW{} around assistants), holds when
search events are added back to the assistant side (\WebBeforeUW{} versus \WebAfterUW{}),
and holds \emph{within} panelists contributing both session types: the per-user
before-minus-after content gap is \WithinUserAIDirection{} percentage points around
assistants and \WithinUserSearchDirection{} around search. The single cleanest recomposition
estimand is the paired difference of these directions---(assistant before minus after) minus
(search before minus after), bootstrapped as one quantity per user---which is
\DirectionalRecomposition{} percentage points. Descriptively, the search span
more often precedes its content, while assistant spans more often follow observed web
activity. Search-centered sessions also
interleave more (bridge \SearchTopoBridgeUW{} versus \TopoBridgeUW{}): the query-click-query
loop is a more tightly alternating structure than dialogue, which concentrates consecutive
assistant events. This is the observed positional grammar of the two anchors.

\paragraph{Within the same user.} Because assistant adopters may simply browse differently,
the cleanest contrast conditions on the person. Among panelists contributing both session
types in the month, the within-user difference in contained shares (assistant-centered minus
search-centered, equal-user weighting) is \WithinUserContainedDiff{} percentage points.
This conditions on the person, not on what the person brought to each surface; the task
mix is what the next construction takes up.

\paragraph{Robust to observable destination-domain strata.} Because each anchor receives its own mix of needs, the
reversal could in principle be composition: assistants attracting the kinds of tasks whose
content naturally precedes them. Recomputing the paired difference-of-directions within
coarse task categories gives that reading no support within the strata the trace can resolve (Table~\ref{tab:taskmix}). Categories
are assigned from each session's content-pageview domains against head-domain
lexicons fixed before this analysis (the shopping, news, reference, coding, and leisure
host lists shared across our panel studies); \TaskCoverageAI{} of assistant-anchored and \TaskCoverageSearch{} of
search-anchored sessions carry a category, and sessions with no categorizable content
pageview fall in ``uncategorized,'' which by construction includes every
assistant-contained session. The contrast is positive with intervals excluding zero in all
six strata; coding/documentation carries the widest interval and excludes zero too. Pooled
over the categorized sessions alone it is \TaskPooledCategorizedDiff{} percentage points,
so the result is not carried by the uncategorized bucket. Both components flip sign within
every stratum, assistant before-minus-after positive and search before-minus-after
negative. These categories are read from the same content-pageview stream that defines the
before-and-after outcome, so this is destination-domain robustness rather than an
outcome-independent task label; whether the reversal is a task-mix artifact on the semantic
task and task-stage dimensions the trace cannot see (intent, urgency, stage) is the question
the human-validated task-matching study is designed to settle.

\begin{table}[t]
\caption{The reversal within coarse destination-domain categories: paired within-user
difference-of-directions, assistant minus search, in percentage points with 95\%
user-clustered intervals. ``Uncategorized'' covers sessions with no categorizable content
pageview, including all assistant-contained sessions.}
\label{tab:taskmix}
\small
\begin{tabular}{lr}
\toprule
category & paired difference (pp) \\
\midrule
shopping & \TaskShoppingDiff \\
news & \TaskNewsDiff \\
coding / documentation & \TaskCodingDiff \\
reference & \TaskReferenceDiff \\
leisure & \TaskLeisureDiff \\
uncategorized & \TaskOtherDiff \\
\bottomrule
\end{tabular}
\end{table}

\paragraph{Composition, not just position.} The shapes also differ in what fills them.
Composition is summarized session-weighted: within-session event shares and switch counts
are averaged over pooled sessions with user-clustered intervals (surfaces are assistant,
search, and content; a switch is an adjacent surface change in the time-ordered stream;
durations are pooled medians). Within assistant-centered sessions, assistant events make up
\CompAIAssistantShare{} of
steps, search events \CompAISearchShare{}, and content pageviews \CompAIContentShare{}; the
median session lasts \CompAIDurationMin{} minutes and switches surface
\CompAISwitchesCI{} times on average. Search-centered sessions are \CompSearchSearchShare{}
search and \CompSearchContentShare{} content by steps, with a median duration of
\CompSearchDurationMin{} minutes and \CompSearchSwitchesCI{} surface switches. These are
event-share summaries of observed activity, not time-use estimates; dwell is observed only
for pageviews.

\paragraph{Stability within the window.} Splitting the month by market and by assistant
(single-assistant sessions; Perplexity is too thin to split; the search comparator is
attributed to each user's most-frequent provider) moves the \emph{levels}
by tens of points while preserving the containment ordering. Assistant-side containment, user-weighted throughout this
paragraph, is
\MarketUSAIContained{} in the United States and \MarketGBAIContained{} in Great Britain,
and \AsstGeminiContained{} for Gemini against \AsstChatGPTContained{} for ChatGPT---levels
are clearly population- and provider-dependent. In every split, assistant-side
containment exceeds the search comparator's (\MarketUSSearchContained{} and
\MarketGBSearchContained{} in the two markets). The directional mirror, with user-clustered
intervals, holds in every split: the search side is strongly negative throughout
(\MarketUSSearchDirection{} and \MarketGBSearchDirection{} by market;
\AsstChatGPTSearchDirection{} and \AsstGeminiSearchDirection{} by provider), and the
assistant side stays positive with intervals excluding zero (\MarketUSAIDirection{} and
\MarketGBAIDirection{} by market; \AsstChatGPTDirection{} for ChatGPT). The one thin case is
the Gemini assistant side (\AsstGeminiDirection{}): still positive but small, with a lower
bound close to zero, so the mirror's assistant-side \emph{magnitude} is provider-dependent
even though its sign and the containment ordering are not. The ordering travels; the exact
shares, and the directional magnitude, are contextual.

\paragraph{Replication in a second month.} We repeat the entire construction on a second,
adjacent month of the same panel, March 2026, changing nothing else. Every headline
replicates. Assistant-centered containment is \CrossMarAIContained{}
user-weighted against February's \CrossFebAIContained{}; the search comparator is
\CrossMarSearchContained{} against \CrossFebSearchContained{}; bridge is \CrossMarAIBridge{}
(assistant) and \CrossMarSearchBridge{} (search). The within-user contained contrast is
\CrossMarWithinContained{} percentage points in March against \CrossFebWithinContained{} in
February, and the paired directional recomposition is \CrossMarDirectional{} against
\CrossFebDirectional{}; the estimates are consistent across the two months. The exact shares are
context-dependent (they move by tens of points across markets and providers within a
month), but the two structural results---assistant use is far more contained than search,
and the before/after order reverses between the anchors---hold unchanged across the two
months we can observe.

\paragraph{Comparator selection.} The two session populations are selected by
their own anchors: sessions enter the AI-centered frame by containing assistant activity and
the comparator by containing search without assistants, so the comparison describes how the
two anchors organize their surroundings, not what would happen if one replaced the other.
Google's search-embedded AI is excluded from the assistant stream by construction, which
means some comparator searches were themselves AI-answered on the results page. We expect
this to raise search-side containment (an embedded answer needs no click) and therefore to
shrink the reported containment gap, making the contrast conservative in that respect; but
embedded answers could shift other positions in either direction, so we state this as an
expectation, not a demonstrated bound.

\paragraph{All-search sensitivity.} Excluding assistant-containing sessions from the
comparator is a selection choice, so we recompute it keeping \emph{every} search-containing
session (including those that also hold assistant events), with search as the anchor and both
content and assistant events as non-anchor activity. Search-contained is
\AllSearchContainedUW{} user-weighted and bridge \AllSearchBridgeUW{}, close to the
assistant-excluded comparator (\SearchTopoContainedUW{} and \SearchTopoBridgeUW{}). The old/new
containment gap is therefore not an artifact of dropping mixed AI-and-search sessions; those
sessions are a small minority and organize their content much as pure search sessions do.

\section{Coexistence, Not Displacement}
\label{sec:coexistence}
It is tempting to read a large contained share as evidence that AI displaces search.
(Appendix~\ref{sec:permeability} reports how quickly a next observed web event of any kind
follows an assistant session---\NextWebOneDay{} within a day---under a strict risk-set rule;
those curves count unrelated browsing and are horizon diagnostics, not task-continuation
estimates.
The thirty-minute point of that curve is not independent evidence: with a thirty-minute session
gap, any web event that soon is by construction inside the same session, so it restates the
after-marginal \WebAfterEW{} rather than adding to it.) Our data do not
license that verdict, and we do not make it. Whether AI creates or destroys search demand is a
counterfactual that observational cross-surface data cannot identify against burst-selected
adoption~\cite{barabasi2005bursts}; the methods companion proves that a single-surface event window
does not separate the two without further assumptions~\cite{iannelli2026burst}. What we can say is
compositional and holds without a counterfactual: containment and cross-surface coexistence occur at the same time. About a
third of sessions are assistant-contained user-weighted, while search appears somewhere in just under
half. Search is observed before the first assistant event more often than after the last, but neither
position establishes substitution, verification, or causation. We state this as coexistence precisely
because ``AI does not displace search'' is a demand counterfactual the design cannot identify.

\section{Discussion}
The central result is not the contained share by itself. It is the reversal in direction.
In the same panelists, conventional search more often precedes content, while assistant use
more often follows it. The paired difference is \DirectionalRecomposition{} percentage
points and repeats in a second month. This is the observed meaning of the difference:
assistant use occupies a different position relative to search and browsing than a
conventional query does. We report observed position, not a counterfactual about what the
journey would otherwise have contained.
Table~\ref{tab:claims} states what each headline claim rests on and what observation would
overturn it.

\begin{table}[t]
\caption{Claims and constraints: what licenses each claim, and what would overturn it.}
\label{tab:claims}
\small\setlength{\tabcolsep}{4pt}\renewcommand{\arraystretch}{1.1}
\begin{tabular}{>{\raggedright\arraybackslash}p{2.15cm}
                >{\raggedright\arraybackslash}p{2.65cm}
                >{\raggedright\arraybackslash}p{2.65cm}}
\toprule
claim & rests on & would overturn it \\
\midrule
no single position describes assistant use &
direct classification of every AI-containing session; gap and coverage sensitivity &
a defensible gap or coverage rule concentrating most sessions in one class (max observed:
\GapFifteenContained{} contained at 15 min) \\
\addlinespace[5pt]
assistant sessions more contained than search sessions (within user) &
identical construction on both anchors; within-user contrast &
a search definition reversing the ordering (none of four definitions does) \\
\addlinespace[5pt]
directional reversal (content after search, before assistants) &
content-only on both sides; paired within-user difference; second-month replication;
category-stratified sensitivity &
a market, provider, month, or task-category split flipping the sign (none observed; the
Gemini magnitude is small) \\
\addlinespace[5pt]
recomposition, not displacement &
descriptive design; anchors select their own sessions &
outside the design by construction; a demand verdict needs task matching or exogenous
variation \\
\bottomrule
\end{tabular}
\end{table}

The result suggests a three-layer view of the emerging information environment. Search
remains an access layer that commonly opens movement toward documents. Browsing remains an
evidence and discovery layer. Conversational AI can become a synthesis layer deeper in the
journey, while also sometimes opening it, containing it, or participating in repeated
cross-surface movement. These are conceptual functions motivated by the aggregate temporal
grammar, not labels assigned to individual sessions. Establishing them session by session
requires semantic task validation.

The topology also changes what should count as activity. A multi-response assistant
session is not an empty interval merely because no pageview follows it. Conversely, the
absence of an external event does not establish that a need was resolved. Weighting,
inactivity gaps, and browsing-observation density all move the apparent balance. The durable
findings are therefore the plurality of positions, the old/new directional contrast, and
the need to keep observed containment separate from inferred task completion.

\paragraph{Implications.} Retrieval and HCI evaluation should treat prompts and responses
as first-class session events and report the rule linking them to web behavior. Product
evaluation should distinguish assistant-first, assistant-last, contained, and interleaved
contexts before assigning functions to transitions. Ecosystem measurement should separate
three questions that are often collapsed: where activity occurs, how activity is composed,
and whether total demand changes. The first two are observable here; causal displacement is
not.

\paragraph{Boundaries of the claims.} We do not identify whether co-timed events
share one task, whether an assistant resolved the need, or whether assistants caused the
observed difference. We do not separate workbench production from information seeking
because doing so requires prompt-content classification intentionally excluded from the
primary extract. We also do not infer verification from an onward visit or compare
citation-forward and citation-light assistants without adjustment for their different
users, tasks, and interaction styles. These are extensions of the measurement framework,
not conclusions hidden inside the present topology.

\section{Conclusion}
The familiar query-to-content sequence now coexists, in the same panelists, with temporal
sessions in which people bring prior browsing into
dialogue, remain inside dialogue, begin from dialogue, or move repeatedly between the
assistant and the web. Across two months, the old and new anchors show a consistent
directional contrast: search tends to open the observed journey, while conversational
assistants sit deeper inside it.

That is the new shape of search. It is not a shorter version of the old one, and the data
do not establish that assistants caused it. It is a measurable recomposition of where
observed information-seeking activity occurs---and a reason to study dialogue, search, and
browsing as parts of one cross-surface system.

\section*{Ethics and Human-Subjects Statement}
This is secondary analysis of previously collected, opt-in, de-identified web-browsing, search, and
conversational-AI events. Panelists consented to behavioral measurement, including the linkage of
assistant activity with browsing and search on metered devices, under the panel provider's
terms; the authors did not intervene on user experience or contact participants, and no new data were
collected for this study. The authors treated the work as secondary aggregate measurement under the
provider's consent and governance process; it was not submitted for separate institutional ethics
review. The analyses use event timestamps, prompt/response roles, hashed
provider-session identifiers, sequence, content length, and coarse domains, with no page content or conversation
text. Results are reported only in aggregate, as rates, ratios, and
user-clustered intervals; no raw conversations, searches, individual URLs, or sample-size statistics
that could expose panel composition are released. One contextual point deserves naming: linking AI use
to browsing and search for the same user is more sensitive than either stream alone, and our use of
that linkage is confined to aggregate measurement. Because conversational content can carry medical,
financial, or otherwise sensitive material, we restrict this study to timestamp-and-domain features and
keep content out of the analyzed streams entirely.

\section{Limitations}
\label{sec:limits}
First, \emph{scope of claim}: we characterize temporal-session composition and make no causal-volume
claim. The topology does not identify demand created or destroyed, which an observational panel cannot
recover against burst-selection~\cite{barabasi2005bursts,iannelli2026burst}. Second,
\emph{task identity}: temporal proximity does not establish semantic continuity. Concurrent needs can merge, one need can split, and
provider sessions are not validated task labels. Two construction details cut the same way and are worth naming. Assistant hosts outside the
canonical list (for example Grok, DeepSeek, Meta AI, you.com) are not recognised as assistant
surfaces, so their pageviews stay in the web stream and count as ordinary web activity; a session
that moves from a recognised assistant to an unrecognised one is therefore read as AI-last or
bridge rather than contained. And the extract admits assistant events whose captured content is
between 2 and 30{,}000 bytes, so an unusually long response is dropped from the stream, which can
move the last observed assistant timestamp earlier and reclassify a following pageview from
``between'' to ``after''. Both push the same way -- against containment and against the
directional contrast -- so the reported figures are the conservative side of each.
Unmetered assistant channels (native mobile apps, API access) are unobserved; to the
extent quick mobile lookups skew toward contained or AI-first use, those shares are
understated here, while the directional contrast, which conditions on observed web
activity around the span, is less exposed though not immune. Third, \emph{observation density}: the primary frame
requires at least one pageview-active day, not guaranteed continuous instrumentation. The strong movement
across active-day thresholds shows that containment is partly a measurement-density quantity; the
threshold populations also differ behaviorally, so no row is an unbiased correction for another.
Fourth, \emph{session boundaries}: from a 15- to a 60-minute gap the contained share falls from
\GapFifteenContained{} to \GapSixtyContained{} and bridge rises from \GapFifteenBridge{} to
\GapSixtyBridge{}, while the one-sided classes move by at most a point. Fifth, \emph{deferred constructs}: workbench-versus-seeking,
same-task continuity, resolution, internalization, verification, and satisfaction require a separately
governed content-validation sample. Sixth, \emph{measurement}: domains are coarse; onward visits are
timing-linked rather than click-observed; and next-event curves count unrelated activity.
Seventh, \emph{comparator selection}: the search-centered comparator is selected by its own
anchor, so old-versus-new contrasts are descriptive of how each anchor organizes its
surroundings; the within-user contrast conditions on the person, and the category-stratified
sensitivity conditions only coarsely on the task mix, so semantic task matching remains
open. Finally,
\emph{external validity}: the study covers the United States and Great Britain, an AI-forward opt-in
panel, two adjacent months (February and March 2026; the structural results replicate across
them, but two months is still a short and possibly atypical window), observed visits rather
than reading depth, and standalone assistants only. All reported
shares carry user-clustered intervals, but precision does not remove these construct limitations.

\bibliographystyle{ACM-Reference-Format}
\bibliography{references}

@article{barabasi2005bursts,
	title = {The origin of bursts and heavy tails in human dynamics},
	volume = {435},
	doi = {10.1038/nature03459},
	journal = {Nature},
	author = {Barab\'{a}si, Albert-L\'{a}szl\'{o}},
	year = {2005},
	pages = {207--211},
}

@misc{semrush2025expansion,
	title = {{ChatGPT} {Is} {Not} {Replacing} {Google}: {It}'s {Expanding} {Search}},
	url = {https://www.semrush.com/blog/google-usage-after-chatgpt-adoption/},
	author = {{Semrush}},
	year = {2025},
	note = {Status: Semrush blog/report; analysis of 260 billion rows of clickstream data, Jan 2024--Jun 2025, same-user before/after ChatGPT adoption},
}

@inproceedings{yu2026searchable,
	title = {From {Searchable} to {Non}-{Searchable}: {Generative} {AI} and {Information} {Diversity} in {Online} {Information} {Seeking}},
	author = {Yu, Yulin and Li, Yizhou and Suri, Siddharth and Counts, Scott},
	year = {2026},
	booktitle = {Proceedings of the Extended Abstracts of the 2026 {CHI} Conference on Human Factors in Computing Systems ({CHI} {EA} '26)},
	publisher = {ACM},
	address = {New York, NY, USA},
	pages = {1--6},
	doi = {10.1145/3772363.3798802},
	note = {Status: {CHI} {EA} '26 Extended Abstracts},
}

@misc{zhao2024wildchat,
	title = {{WildChat}: {1M} {ChatGPT} {Interaction} {Logs} in the {Wild}},
	author = {Zhao, Wenting and Ren, Xiang and Hessel, Jack and Cardie, Claire and Choi, Yejin and Deng, Yuntian},
	year = {2024},
	note = {Status: International Conference on Learning Representations (ICLR)},
}

@article{pirolli1999foraging,
	title = {Information {Foraging}},
	volume = {106},
	number = {4},
	journal = {Psychological Review},
	author = {Pirolli, Peter and Card, Stuart},
	year = {1999},
	pages = {643--675},
}

@article{marchionini2006exploratory,
	title = {Exploratory search: from finding to understanding},
	volume = {49},
	number = {4},
	journal = {Communications of the ACM},
	author = {Marchionini, Gary},
	year = {2006},
	pages = {41--46},
}

@article{liu2024lost,
	address = {Cambridge, MA},
	title = {Lost in the {Middle}: {How} {Language} {Models} {Use} {Long} {Contexts}},
	volume = {12},
	doi = {10.1162/tacl_a_00638},
	journal = {Transactions of the {Association} for {Computational} {Linguistics}},
	publisher = {MIT Press},
	author = {Liu, Nelson F. and Lin, Kevin and Hewitt, John and Paranjape, Ashwin and Bevilacqua, Michele and Petroni, Fabio and Liang, Percy},
	year = {2024},
	pages = {157--173},
}

@misc{iannelli2026fptp,
	title = {{From} {Prompt} to {Purchase}: {How} {AI} {Brand} {Recommendations} {Move} {Consumers} on the {Open} {Web}},
	url = {https://arxiv.org/abs/2606.10907},
	author = {Iannelli, Michael and Ai, Alan},
	year = {2026},
	note = {Accepted at the 5th Workshop on End-to-End Customer Journey Optimization (KDD 2026); to appear},
}

@misc{iannelli2026burst,
	title = {Event-{Time} {Confounding} {Under} {Bursty} {Human} {Dynamics}},
	url = {https://arxiv.org/abs/2608.21294},
	author = {Iannelli, Michael and Ai, Alan},
	year = {2026},
	note = {arXiv:2608.21294},
}

@misc{huang2026answerbubbles,
	title = {Answer {Bubbles}: {Information} {Exposure} in {AI}-{Mediated} {Search}},
	author = {Huang, Michelle and Goyal, Agam and Saha, Koustuv and Chandrasekharan, Eshwar},
	year = {2026},
	note = {Status: arXiv:2603.16138},
}

@techreport{chatterji2025chatgpt,
	title = {How {People} {Use} {ChatGPT}},
	institution = {National Bureau of Economic Research},
	author = {Chatterji, Aaron and Cunningham, Tom and Deming, David and Hitzig, Zo{\"e} and Ong, Christopher and Shan, Carl and Wadman, Kevin},
	year = {2025},
}

@misc{aral2026aisearch,
	title = {The {Rise} of {AI} {Search}: {Implications} for {Information} {Markets} and {Human} {Judgement} at {Scale}},
	author = {Aral, Sinan and Li, Haiwen and Zuo, Rui},
	year = {2026},
	note = {Status: arXiv:2602.13415},
}

@article{kuhlthau1991, author={Kuhlthau, Carol C.}, title={Inside the search process: Information seeking from the user's perspective}, journal={Journal of the American Society for Information Science}, volume={42}, number={5}, pages={361--371}, year={1991}}

@book{kuhlthau2004, author={Kuhlthau, Carol C.}, title={Seeking Meaning: A Process Approach to Library and Information Services}, edition={2}, publisher={Libraries Unlimited}, address={Westport, CT, USA}, year={2004}}

@article{belkin1982ask, author={Belkin, Nicholas J. and Oddy, Robert N. and Brooks, Helen M.}, title={ASK for information retrieval: Part I. Background and theory}, journal={Journal of Documentation}, volume={38}, number={2}, pages={61--71}, year={1982}, doi={10.1108/eb026722}}

@article{dervin1998sensemaking, author={Dervin, Brenda}, title={Sense-making theory and practice: an overview of user interests in knowledge seeking and use}, journal={Journal of Knowledge Management}, volume={2}, number={2}, pages={36--46}, year={1998}, doi={10.1108/13673279810249369}}

@article{wilson1999models, author={Wilson, T. D.}, title={Models in information behaviour research}, journal={Journal of Documentation}, volume={55}, number={3}, pages={249--270}, year={1999}, doi={10.1108/EUM0000000007145}}

@article{rieh2016learning, author={Rieh, Soo Young and Collins-Thompson, Kevyn and Hansen, Preben and Lee, Hye-Jung}, title={Towards searching as a learning process: A review of current perspectives and future directions}, journal={Journal of Information Science}, volume={42}, number={1}, pages={19--34}, year={2016}, doi={10.1177/0165551515615841}}

@article{vakkari2016learning, author={Vakkari, Pertti}, title={Searching as learning: A systematization based on literature}, journal={Journal of Information Science}, volume={42}, number={1}, pages={7--18}, year={2016}, doi={10.1177/0165551515615833}}

@article{sparrow2011google, author={Sparrow, Betsy and Liu, Jenny and Wegner, Daniel M.}, title={Google effects on memory: Cognitive consequences of having information at our fingertips}, journal={Science}, volume={333}, number={6043}, pages={776--778}, year={2011}, doi={10.1126/science.1207745}}

@inproceedings{radlinski2017framework, author={Radlinski, Filip and Craswell, Nick}, title={A Theoretical Framework for Conversational Search}, booktitle={Proceedings of the 2017 Conference on Conference Human Information Interaction and Retrieval (CHIIR)}, publisher={ACM}, address={New York, NY, USA}, pages={117--126}, year={2017}, doi={10.1145/3020165.3020183}}

@inproceedings{huang2009reformulation, author={Huang, Jeff and Efthimiadis, Efthimis N.}, title={Analyzing and evaluating query reformulation strategies in web search logs}, booktitle={Proceedings of the 18th ACM Conference on Information and Knowledge Management (CIKM)}, publisher={ACM}, address={New York, NY, USA}, pages={77--86}, year={2009}, doi={10.1145/1645953.1645966}}

@article{belkin1995cases, author={Belkin, Nicholas J. and Cool, Colleen and Stein, Adelheit and Thiel, Ulrich}, title={Cases, scripts, and information-seeking strategies: On the design of interactive information retrieval systems}, journal={Expert Systems with Applications}, volume={9}, number={3}, pages={379--395}, year={1995}, doi={10.1016/0957-4174(95)00011-W}}

@inproceedings{jarvelin2008session, author={J{\"a}rvelin, Kalervo and Price, Susan L. and Delcambre, Lois M. L. and Nielsen, Marianne Lykke}, title={Discounted Cumulated Gain Based Evaluation of Multiple-Query IR Sessions}, booktitle={Advances in Information Retrieval (ECIR)}, publisher={Springer}, address={Berlin, Heidelberg}, pages={4--15}, year={2008}, doi={10.1007/978-3-540-78646-7_4}}

@article{catledge1995characterizing, author={Catledge, Lara D. and Pitkow, James E.}, title={Characterizing browsing strategies in the {World-Wide Web}}, journal={Computer Networks and ISDN Systems}, volume={27}, number={6}, pages={1065--1073}, year={1995}, doi={10.1016/0169-7552(95)00043-7}}

@article{broder2002taxonomy, author={Broder, Andrei}, title={A Taxonomy of Web Search}, journal={SIGIR Forum}, volume={36}, number={2}, pages={3--10}, year={2002}, publisher={Association for Computing Machinery}, doi={10.1145/792550.792552}}

@inproceedings{lichtenegger2026taxonomy, author={Lichtenegger, Elsa and Urman, Aleksandra and Hannak, Aniko}, title={A New Taxonomy of Web Search: A User-Centered Framework for Search Intent in the AI Era}, booktitle={Proceedings of the 2026 CHI Conference on Human Factors in Computing Systems}, series={CHI '26}, articleno={692}, numpages={16}, year={2026}, publisher={Association for Computing Machinery}, address={New York, NY, USA}, doi={10.1145/3772318.3791050}}

@article{zamani2023cis, author={Zamani, Hamed and Trippas, Johanne R. and Dalton, Jeff and Radlinski, Filip}, title={Conversational Information Seeking}, journal={Foundations and Trends in Information Retrieval}, volume={17}, number={3--4}, pages={244--456}, year={2023}, publisher={Now Publishers}, doi={10.1561/1500000081}}

@inproceedings{shah2022situating, author={Shah, Chirag and Bender, Emily M.}, title={Situating Search}, booktitle={Proceedings of the 2022 Conference on Human Information Interaction and Retrieval}, series={CHIIR '22}, pages={221--232}, year={2022}, publisher={Association for Computing Machinery}, address={New York, NY, USA}, doi={10.1145/3498366.3505816}}

@misc{pew2025aioverviews, author={{Pew Research Center}}, title={Google Users Are Less Likely to Click on Links When an {AI} Summary Appears in the Results}, howpublished={Pew Research Center, Short Reads}, month=jul, year={2025}, note={Published July 22, 2025}, url={https://www.pewresearch.org/short-reads/2025/07/22/google-users-are-less-likely-to-click-on-links-when-an-ai-summary-appears-in-the-results/}}

\clearpage
\appendix
\section{Time to the Next Observed Web Event}
\label{sec:permeability}
For each AI-containing temporal session we compute elapsed time from its last assistant event to the
user's next observed web event of any kind. The estimator is a raw cumulative fraction under a strict
risk-set rule for the administrative window end (2026-03-01): a session enters the estimate at horizon
$h$ only if the window extends at least $h$ past its last assistant event, so sessions near the end of
February drop out of long horizons rather than biasing them, and repeated sessions per user enter
individually with uncertainty clustered by user. This is deliberately named by
what it measures: it is not a task-continuation curve because the next event may be unrelated.

\begin{figure}[t]
\centering
\includegraphics[width=0.86\columnwidth]{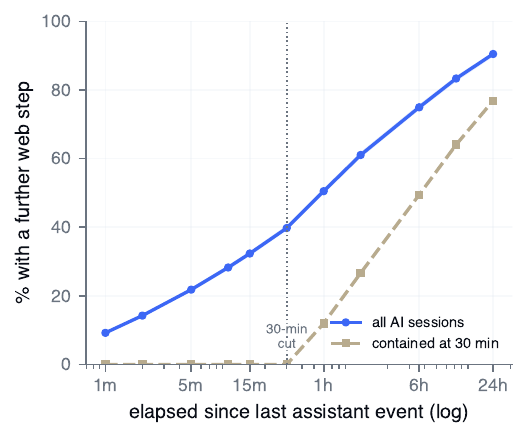}
\caption{Fraction of AI-containing temporal sessions followed by any observed web event, by elapsed
time since the last assistant event (log axis; administrative right-censoring; 95\% user-clustered band
on the all-session curve). The contained-at-30-minutes curve is zero through 30 minutes by construction.
Later events are not asserted to share a task.}
\Description{A line chart with a logarithmic time axis from one minute to 24 hours. The all-session
curve rises from about 9 percent at one minute to 39.7 percent at 30 minutes, 50.5 percent at one hour,
74.9 percent at six hours, and 90.5 percent at 24 hours. The contained-at-30-minutes curve rises after
the threshold to 12.0 percent at one hour, 49.5 percent at six hours, and 76.9 percent at 24 hours.}
\label{fig:continuation}
\end{figure}

The reported estimate rises steadily through the conventional cutoff rather than breaking at it: a
next web event appears by five minutes for
\NextWebFiveMinCI{} of sessions, by fifteen minutes for \NextWebFifteenMinCI{}, by thirty minutes for
\NextWebThirtyMinCI{}, and by one hour for \NextWebOneHourCI{}; the analysis does not establish a
natural session boundary at 30 minutes. Among sessions classified contained at
30 minutes, \ContainedNextWebOneHourCI{} have a web event by one hour and
\ContainedNextWebSixHourCI{} by six hours. Across all AI-containing sessions (not the contained
subset), search is less frequent: \NextSearchThirtyMinCI{} have a next
search event by 30 minutes and \NextSearchOneHourCI{} by one hour. These curves quantify sensitivity to
the observation horizon. Their long-horizon values mostly reflect ordinary browsing opportunity and
must not be read as delayed completion of the same need.

\end{document}